\documentclass[pre,twocolumn,superscriptaddress]{revtex4}  

\usepackage[pdftex]{graphicx}
\usepackage{amsmath,amsfonts,amssymb}
\usepackage{morefloats}

\usepackage{hyperref}
\usepackage{xcolor} 

\definecolor{bluemoi}{rgb}{0.25,0.50 ,0.75} 

\hypersetup{
  bookmarksopen=false,
  pdftitle=" ",
  pdfauthor=" ", 
  pdfsubject=" ", 
  pdftoolbar=true, 
  pdfmenubar=true, 
  pdfhighlight=/O, 
  colorlinks=true, 
  pdfpagemode=UseNone, 
  pdfpagelayout=SinglePage, 
  pdffitwindow=true, 
  linkcolor=bluemoi, 
  citecolor=bluemoi, 
  urlcolor=bluemoi 
}

\makeatletter
\renewcommand{\figurename}{Figure}
\makeatother
\makeatletter
\renewcommand{\fnum@figure}{\small\textbf{\figurename~\thefigure}}
\makeatother

\begin{document}

\title{Cross-checking different sources of mobility information}

\author{Maxime Lenormand}\affiliation{Instituto de F\'isica Interdisciplinar y Sistemas Complejos IFISC (CSIC-UIB), Campus UIB, 07122 Palma de Mallorca, Spain}

\author{Miguel Picornell}\affiliation{Nommon Solutions and Technologies, calle Ca\~nas 8, 28043 Madrid, Spain}

\author{Oliva G. Cant\'u-Ros}\affiliation{Nommon Solutions and Technologies, calle Ca\~nas 8, 28043 Madrid, Spain}

\author{Ant{\`o}nia Tugores}\affiliation{Instituto de F\'isica Interdisciplinar y Sistemas Complejos IFISC (CSIC-UIB), Campus UIB, 07122 Palma de Mallorca, Spain}

\author{Thomas Louail}\affiliation{Institut de Physique Th\'{e}orique, CEA-CNRS (URA 2306), F-91191, 
Gif-sur-Yvette, France}\affiliation{G\'eographie-Cit\'es, CNRS-Paris 1-Paris 7 (UMR 8504), 13 rue du
  four, FR-75006 Paris, France}
  
\author{Ricardo Herranz}\affiliation{Nommon Solutions and Technologies, calle Ca\~nas 8, 28043 Madrid, Spain}

\author{Marc Barthelemy}\affiliation{Institut de Physique Th\'{e}orique, CEA-CNRS (URA 2306), F-91191, 
Gif-sur-Yvette, France}\affiliation{Centre d'Analyse et de Math\'ematique Sociales, EHESS-CNRS (UMR
8557), 190-198 avenue de France, FR-75013 Paris, France}

\author{Enrique Fr\'ias-Martinez}\affiliation{Telef\'onica Research, 28050 Madrid, Spain}

\author{Jos\'e J. Ramasco}\affiliation{Instituto de F\'isica Interdisciplinar y Sistemas Complejos IFISC (CSIC-UIB), Campus UIB, 07122 Palma de Mallorca, Spain}

\begin{abstract} 
The pervasive use of new mobile devices has allowed a better characterization in space and time of human concentrations and mobility in general. Besides its theoretical interest, describing mobility is of great importance for a number of practical applications ranging from the forecast of disease spreading to the design of new spaces in urban environments.  While classical data sources, such as surveys or census, have a limited level of geographical resolution (e.g., districts, municipalities, counties are typically used) or are restricted to generic workdays or weekends, the data coming from mobile devices can be precisely located both in time and space.  Most previous works have used a single data source to study human mobility patterns. Here we perform instead a cross-check analysis by comparing results obtained with data collected from three different sources: Twitter, census and cell phones. The analysis is focused on the urban areas of Barcelona and Madrid, for which data of the three types is available. We assess the correlation between the datasets on different aspects: the spatial distribution of people concentration, the temporal evolution of people density and the mobility patterns of individuals. Our results show that the three data sources are providing comparable information. Even though the representativeness of Twitter geolocated data is lower than that of mobile phone and census data, the correlations between the population density profiles and mobility patterns detected by the three datasets are close to one in a grid with cells of $2\times 2$ and $1\times 1$ square kilometers. This level of correlation supports the feasibility of interchanging the three data sources at the spatio-temporal scales considered.
\end{abstract}

\maketitle

\section{INTRODUCTION}

The strong penetration of ICT tools in the society's daily life is opening new opportunities for the research in socio-technical systems \cite{watts07,lazer09,alex09}. Users' interactions with or through mobile devices get registered allowing a detailed description of social interactions and mobility patterns. The sheer size of these datasets opens the door to a systematic statistical treatment while searching for new information. Some examples include the analysis of the structure of (online) social networks~\cite{Lieben05,onnela07,java07,Huberman2008,krishna08,lewis08, Mislove2008Growth,Eagle2009,ferrara12,Grabowicz2013}, human cognitive limitations~\cite{goncalves11-2}, information diffusion and social contagion~\cite{leskovec09-1,lehmann2012,grabowicz2012,Bakshy2012role,Ugander2012Structural}, the role played by social groups~\cite{grabowicz2012,ferrara12}, language coexistence \cite{Mocanu2013} or even how political movements raise and develop~\cite{borge11,bailon2011,conover13}.  

\begin{figure*}
\centering
\includegraphics[width=12cm]{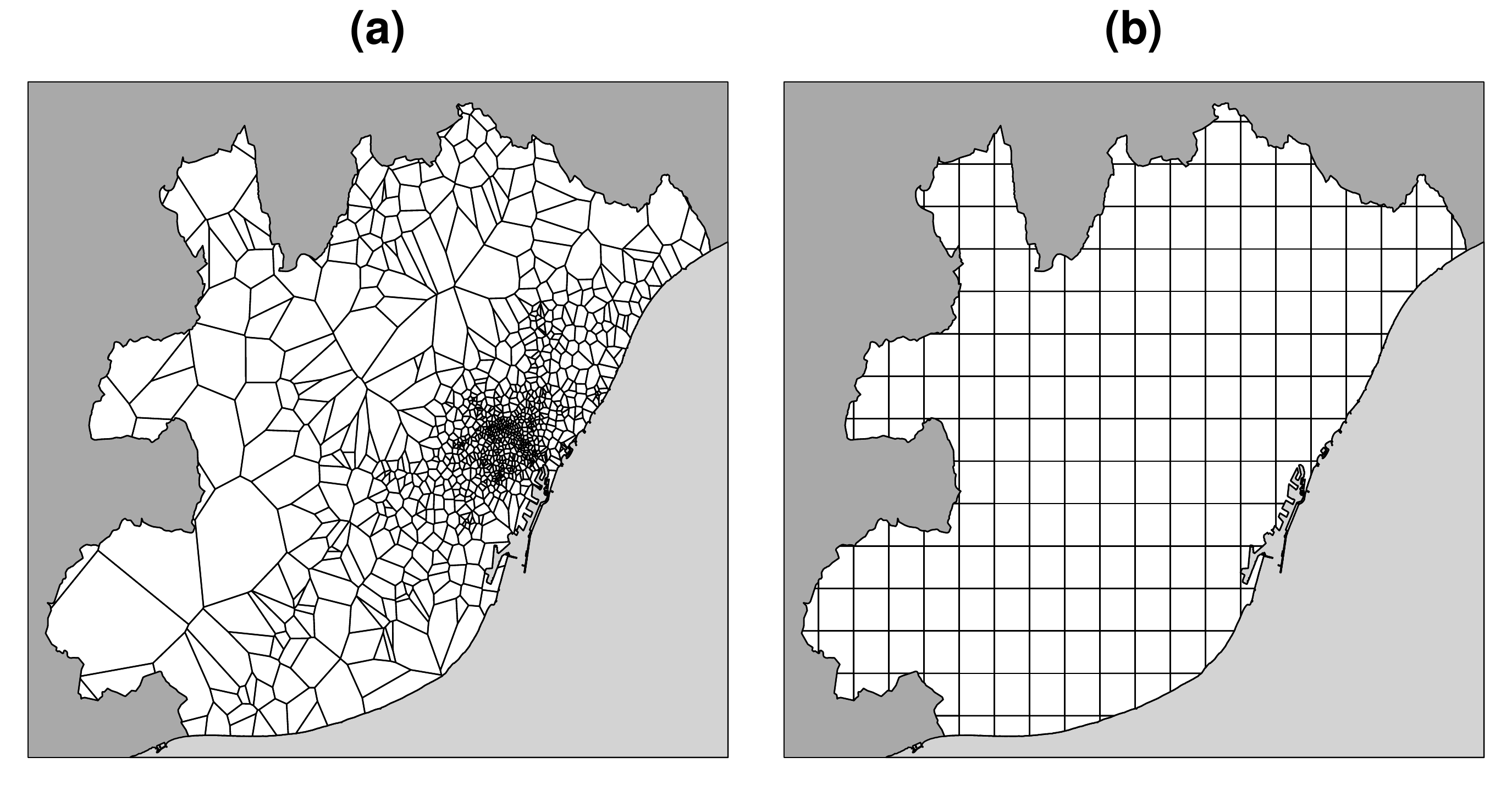}
\caption{Map of the metropolitan area of Barcelona. The white area represents the metropolitan area, the dark grey zones correspond to territory surrounding the metropolitan area and the gray zones to the sea. (a) Voronoi cells around the BTSs. (b) Gird cells of size $2\times2$ $km^2$. \label{Grid}}
\end{figure*}

The analysis of human mobility is another aspect to which the wealth of new data has notably contributed \cite{Brockmann2006,Gonzalez2008,Song10,Bagrow2012,Phithakkitnukoon2012}. Statistical characteristics of mobility patterns have been studied, for instance, in Refs. \cite{Brockmann2006,Gonzalez2008}, finding a heavy-tail decay in the distribution of displacement lengths across users. Most of the trips are short in everyday mobility, but some are extraordinarily long. Besides, the travels are not directed symmetrically in space but show a particular radius of gyration \cite{Gonzalez2008}. The duration of stay in each location also shows a skewed distribution with a few  preferred places clearly ranking on the top of the list, typically corresponding to home and work \cite{Song10}. All the insights gained in mobility, together with realistic data, have been used as proxies for modeling the way in which viruses spread among people \cite{balcan2009} or among electronic devices \cite{wang09}.  Recently, geolocated data has been also used to analyze the structure of urban areas \cite{Ratti2006, Reades2007,Soto2011,Frias2012,Isaacman2012,toole12,pei13,Louail2014}, the relation between different cities \cite{Noulas2012} or even between countries \cite{Hawelka2013}.

Most mobility and urban studies have been performed using data coming essentially from a single data source such as: cell phone data \cite{onnela07,Eagle2009,Gonzalez2008,Song10,Phithakkitnukoon2012,wang09,Ratti2006,Reades2007,Soto2011,Frias2012,Isaacman2012,toole12,pei13,Louail2014}, geolocated tweets \cite{Mocanu2013,borge11,bailon2011,Hawelka2013}, census-like surveys or commercial information \cite{balcan2009}. There is only a few recent exceptions, for instance, epidemic spreading studies \cite{tizzoni13}. When the data has not been “generated” or gathered ad hoc to address a specific question, one fair doubt is how much the results are biased by the data source used. In this work, we compare spatial and temporal population density distributions and mobility patterns in the form of Origin-Destination (OD) matrices obtained from three different data sources for the metropolitan areas of Barcelona and Madrid. This comparison will allow to discern whether or not the results are source dependent. In the first part of the paper the datasets and the methods used to extract the OD tables are described. In the second part of the paper, we present the results. First, a comparison of the spatial distribution of users according to the hour of the day and the day of the week showing that both Twitter and cell phone data are highly correlated on this aspect. Then, we compare the temporal distribution of users by identifying where people are located according to the hour of the day, we show that the temporal distribution patterns obtained with the Twitter and the cell phone datasets are very similar. Finally, we compare the mobility networks (OD matrices) obtained from cell phone data, Twitter and census. We show that it is possible to extract similar patterns from all datasets, keeping always in mind the different resolution limits that each information source may inherently have.

\section{MATERIALS AND METHODS}

This work is focused on two cities: the metropolitan areas of Barcelona \cite{noteB} and Madrid \cite{noteM} both in Spain and for which  data from the three considered sources is available. The metropolitan area of Barcelona contains a population of $3,218,071$ (2009) within an area of $636$ $km^2$. The population of the metropolitan area of Madrid is larger, with $5,512,495$ inhabitants (2009) within an area of $1,935$ $km^2$ \cite{ine}. In order to compare activity and intra mobility in each city, the metropolitan areas are divided into a regular grid of square cells of lateral size $l$ (Figure \ref{Grid}b). Two different sizes of grid cells ($l=1$ $km$ and $l=2$ $km$) are considered in order to evaluate the robustness of the results. Since mobility habits and population concentration may change along the week, we have divided the data into four groups: one, from Monday to Thursday representing a normal working day and three more for Friday, Saturday and Sunday.

The concentration of phone or Twitter users is quantified by defining two three dimensional matrices $T=(T_{g,w,h})$ and $P=(P_{g,w,h})$, accounting, respectively, for the number of Twitter users and the number of mobile phone users in the grid cell $g$ at the hour of the day $h$ and for the group of days $w$. The index for cells $g$ runs in the range $[1,n]$. In the following, details for the three datasets are more thoroughly described. 

\begin{figure*}
\centering
\includegraphics[width=16cm]{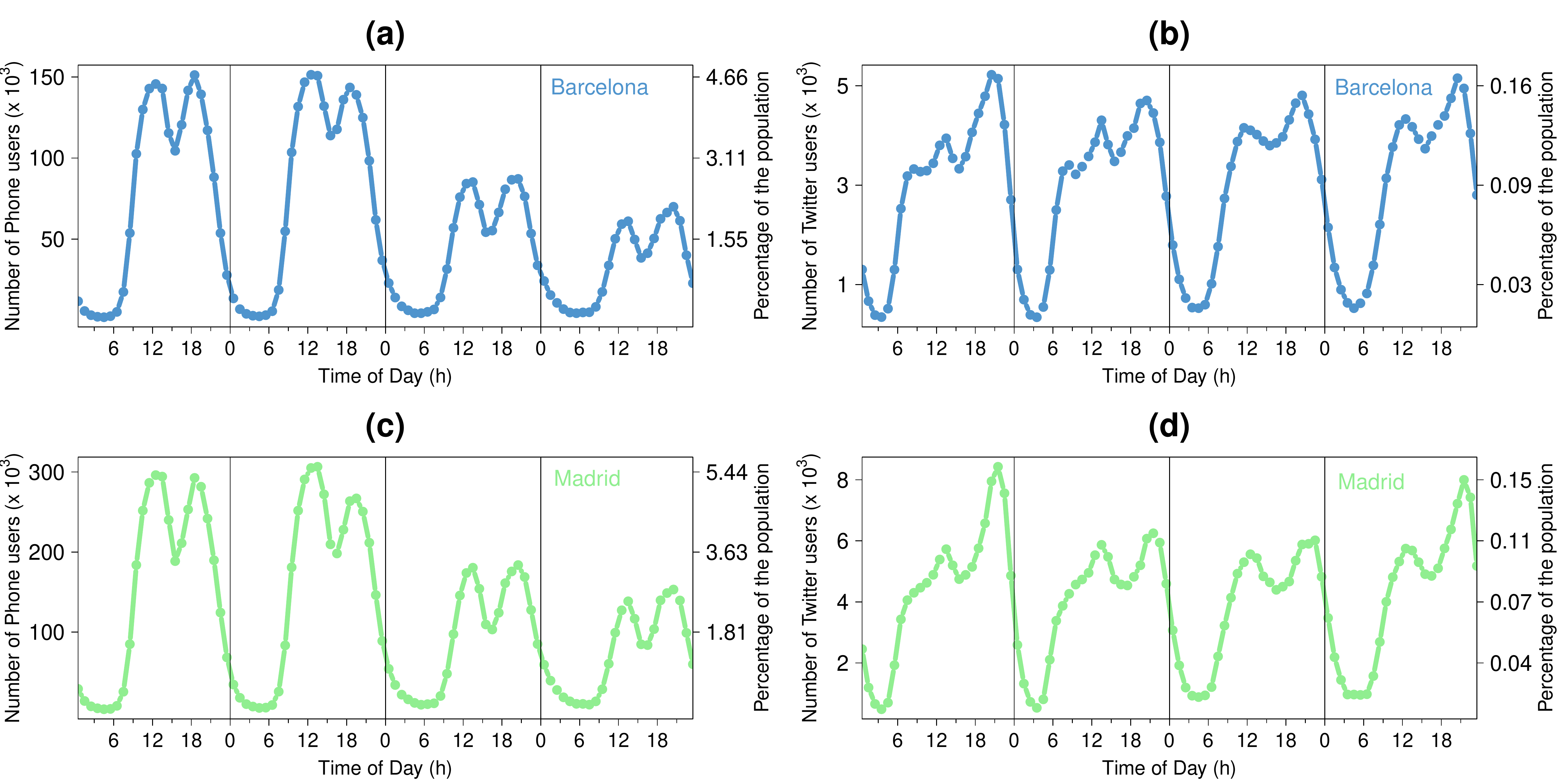}
\caption{Number of mobile phone users per day in Barcelona (a) and Madrid (c) and number of Twitter users in Barcelona (b) and Madrid (d) as a function of the time according to day group $w$. From left to right: weekdays (aggregation from Monday to Thursday), Friday, Saturday and Sunday.\label{NbUsers}}.
\end{figure*}
	
\subsection{Mobile phone data}   

The cell phone data that we are analyzing come from anonymized users' call records collected during 55 days (noted as $D$ hereafter) between September and November 2009. The call records are registered by communication towers (Base Transceiver Station or BTS), identified each by its location coordinates. The area covered by each tower can be approximated by a Voronoi tessellation of the urban areas, as shown in Figure \ref{Grid}a for Barcelona. Each call originated or received by a user and served by a BTS is thus assigned to the corresponding BTS Voronoi area. In order to estimate the number of people in different areas per period of time, we use the following criteria: each person counts only once per hour. If a user is detected in $k$ different positions within a certain 1-hour time period, each registered position will count as ($1/k$) "units of activity". From such aggregated data, activity per zone and per hour is calculated. Consider a generic grid cell $g$ for a day $d$ and hour between $h$ and $h+1$, the $m$ Voronoi areas intersecting $g$ are found and the number of mobile phone users $P_{g,d,h}$ is calculated as follows:
\begin{equation}
P_{g,d,h}=\sum_{v=1}^m N_{v,d,h} \, \frac{\displaystyle A_{v\cap g}}{\displaystyle A_{v}},
\label{Pgdh}
\end{equation}
where $N_{v,d,h}$ is the number of users in a Voronoi cell $v$ on day $d$ at time $h$, $A_{v\cap g}$ is the area of the intersection between $v$ and $g$,  and $A_v$ the area of $v$. The $D$ days available in the database are then divided in four groups according to the classification explained above and the average number of mobile phone users for each day group $w$ is computed as
\begin{equation}
P_{g,w,h}=\frac{\sum_{d\in D_{w}}P_{g,d,h}}{|D_{w}|} .
\label{Pgwh}
\end{equation}

The number of mobile phone users per day for the two the metropolitan areas as a function of the time of day, and according to the day group, are displayed in Figure \ref{NbUsers}. The curves in Figure \ref{NbUsers}a show two peaks, one between noon and $3$pm and another one between $6$pm and $9$pm. They also show that the number of mobile phone users is higher during weekdays than during the weekends. The same curve is obtained for Madrid with about twice the number of users with respect to Barcelona. Further details about the data pre-processing are given in the Appendix (Section \textit{Mobile phone data pre-processing}, Figure S1 and Figure S2).

In order to extract OD matrices from the cell phone calls a subset of users, with a mobility reliably recoverable, was selected. For this analysis we only consider commuting patterns in workdays. The users' home and work are identified as the Voronoi cell most frequently visited on weekdays by each user between $8$ pm and $7$ am (home) and between $9$ am and $5$ pm (work). We assume that there must be a daily travel between home and work location of each individual. Users with calls in more than $40\%$ of the days under study at home or work are considered valid. Aggregating the complete flow over users, an OD commuting matrix is obtained containing in each element the flow of people traveling between a Voronoi cell of residence and another of work. Since the Voronoi areas do not exactly match the grid cells, a transition matrix to change the scale is employed (see Appendix for details).

\begin{figure*}
\centering 
\includegraphics[width=16cm]{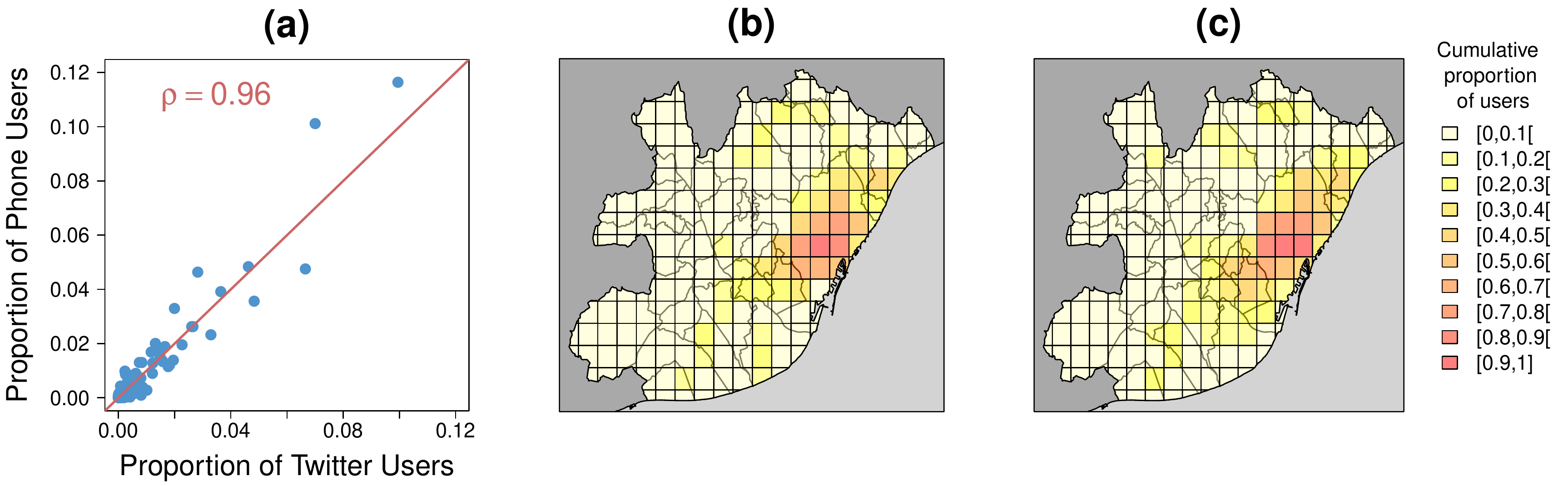}.
\caption{Correlation between the spatial distribution of Twitter users and mobile phone users for the weekdays (aggregation from Monday to Thursday) and from noon to 1pm for the metropolitan area of Barcelona ($l=2$ $km$). (a) Scatter-plot composed by each pair $(T_{g,w,h},P_{g,w,h})$, the values have been normalized (dividing by the total number of users) in order to obtain values between 0 and 1. The red line represents the perfect linear fit  with slope equal to 1 and intercept equal to 0. ((b)-(c)) Spatial distribution of Twitter users (b) and mobile phone users (c). In order to facilitate the comparison of both distributions on the map, the proportion of users in each cell is shown (always bounded in the interval $[0,1]$). \label{Cor}}
\end{figure*}

\subsection{Twitter data}

The dataset comprehends geolocated tweets of $27,707$ users in Barcelona and $50,272$ in Madrid in the time period going from September 2012 to December 2013. These users were selected because it was detected from the general data streaming with the Twitter API \cite{api} that they have emitted at least a geolocated tweet from one of the two cities. Later, as a way to increase the quality of our database, a specific search over their most recent tweets was carried out \cite{antonia13}. As for the cell phone data, the number of Twitter users $T_{g,w,h}$ in each grid cell $g$ per hour $h$ were computed for each day group $w$. The number of Twitter users per day for the metropolitan area of Barcelona according to the hour of the day and the day group is plotted on Figure \ref{NbUsers}b. Analogous to the mobile phone data, this figure shows two peaks, one between noon and 3pm and another one between 6pm and 9pm.  It is worth noting that the mobile phone users represents on average $2\%$ of the total population against $0.1\%$ for the Twitter data. Furthermore, in contrast with the phone users profile curve, the Twitter users' profile curve shows that the number of users does not vary much from weekdays to weekend days. Moreover, we can observe that the number of Twitter users is higher during the second peak than during the first one.

The identification of the OD commuting matrices using Twitter is similar to the one explained for the mobile phones except for two aspects. Since the number of geolocated tweets is much lower than the equivalent in calls per user, the threshold for considering a user valid is set at 100 tweets on weekdays in all the dataset. The other difference is that since the tweets are geolocated with latitude and longitude coordinates, the assignment to the grid cells is done directly without the need of intermediate steps through the Voronoi cells. As for the phone, we keep only users working and living within the metropolitan areas.

\begin{figure*}
\centering
\includegraphics[width=15cm]{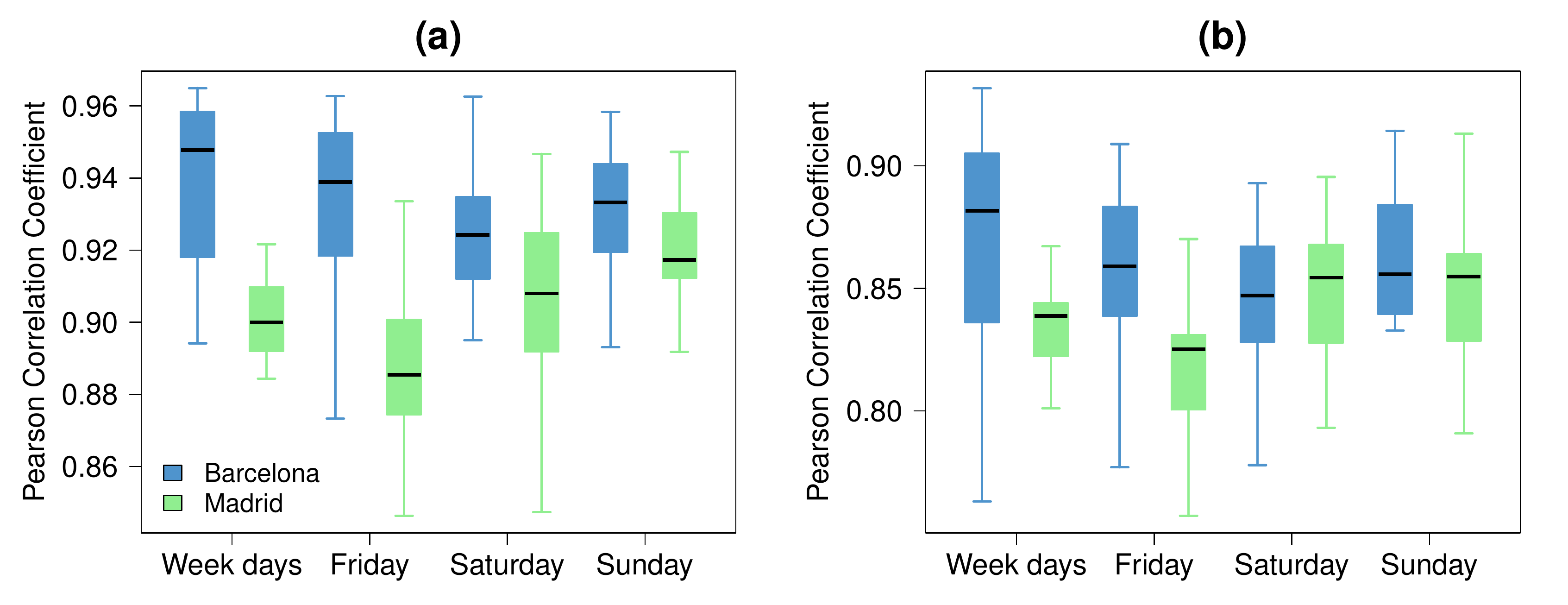}
\caption{Box-plots of the Pearson correlation coefficients obtained for different hours between $T$ and $P$ (from the left to the right: the weekdays (aggregation from Monday to Thursday), Friday, Saturday and Sunday). The blue boxes represent Barcelona. The green boxes represent Madrid. (a) $l=2$ $km$. (b) $l=1$ $km$.\label{box}}
\end{figure*}

\subsection{Census data}

The Spanish census survey of 2011 included a question referring to the municipality of work of each interviewed individual. This survey has been conducted among one fifth of the population. This information, along with the municipality of the household where the interview was carried out, allows for the definition of OD flow matrices at the municipal level \cite{ine}. For privacy reasons, flows with a number of commuters lower than 10 have been removed. The metropolitan area of Barcelona is composed of $36$ municipalities, while the one of Madrid contains $27$ municipalities. In addition to the flows, we have obtained  the GIS files with the border of each municipality from the census office. This information is used to map the OD matrices from Twitter or the cell phone data to this more coarse-grained spatial scale to compare mobility patterns across datasets.

\section{RESULTS}

\subsection{Spatial distribution}

\begin{figure*}
\centering 
\includegraphics[width=16cm]{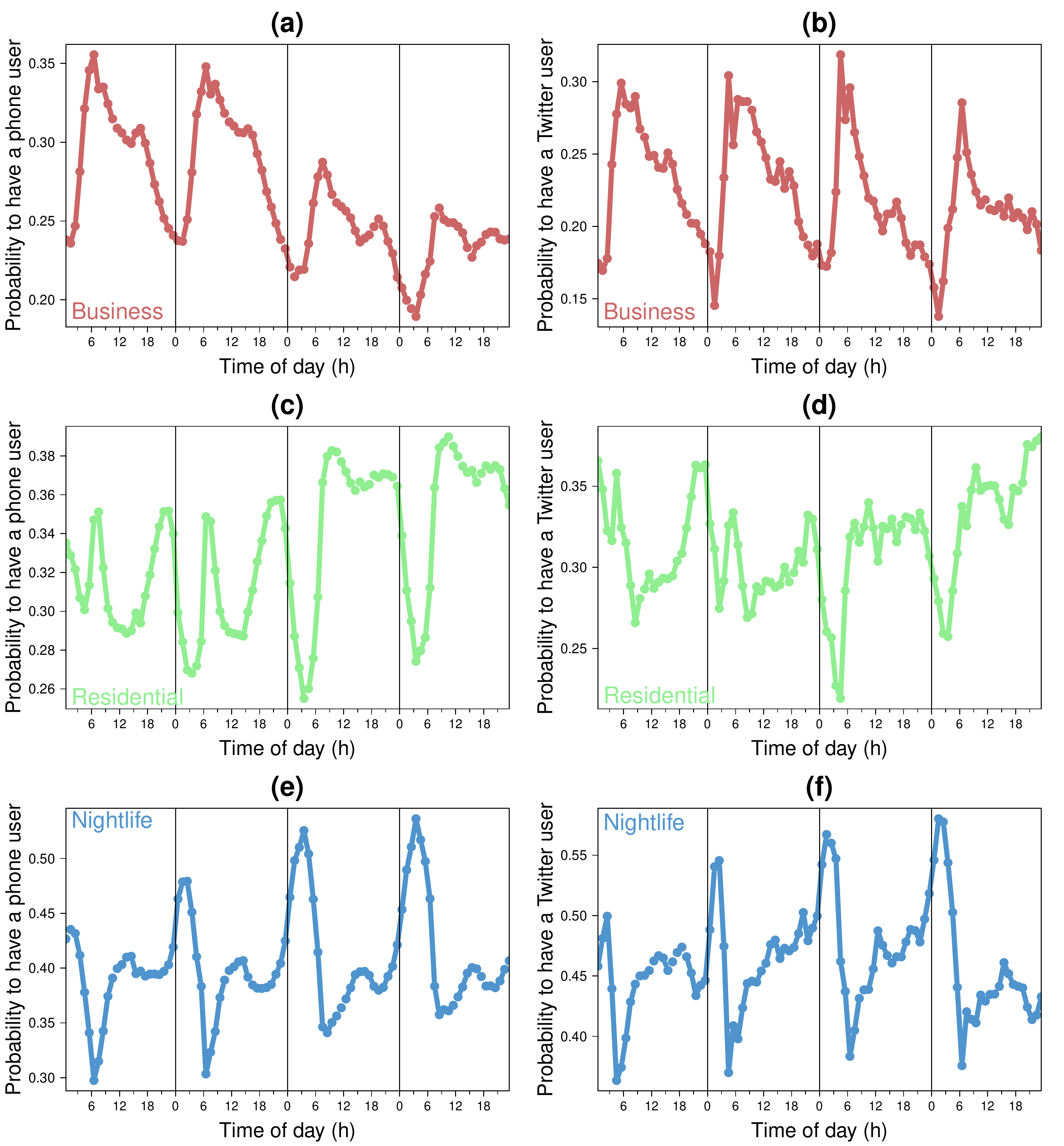}.
\caption{Temporal distribution patterns for the metropolitan area of Barcelona ($l=2$ $km$). (a), (c) and (e) Mobile phone activity; (b), (d) and (f) Twitter activity; (a) and (b) Business cluster; (c) and (d) Residential/leisure cluster; (e) and (f) Nightlife cluster.\label{Patterns}}
\end{figure*}

A first question to address is how much the human activity level is similar or not when estimated from Twitter, $T$, or from cell phone data $P$ across the urban space in grid cells of $2$ by $2$ $km$. To quantify similarity, we start by depicting in Figure \ref{Cor} a scatter plot composed by each pair $(T_{g,w,h},P_{g,w,h})$ for every grid cell of the metropolitan area of Barcelona taking $w$ as  the weekdays (aggregation from Monday to Thursday). The hour $h$ is set from midday to 1pm. A first visual inspection tells us that the agreement between the activity inferred from each dataset is quite good. In fact, the Pearson correlation coefficient between the two estimators of activity is of $\rho = 0.96$. Furthermore, the portion of activity can be depicted on two maps as in Figure \ref{Cor}b and c. The similarity of the areas of concentration of the activity is patent.

More systematically, we plot in Figure \ref{box}a, the box-plots of the Pearson correlation coefficients for each day group and both case studies as observed for different hours. We obtain in average a correlation of $0.93$ for Barcelona and $0.89$ for Madrid. Globally, the correlation coefficients have higher value for Barcelona than for Madrid probably because the metropolitan area of Madrid is about four times larger than the one of Barcelona. It is interesting to note that the average correlation remains high even if we increase the resolution by using a value of $l$ equal to $1$ $km$. Indeed, we obtain in average a correlation of $0.85$ for Barcelona and $0.83$ for Madrid at that new scale (Figure \ref{box}b).

\subsection{Temporal distribution}

After the spatial distribution of activity, we investigate the correlation between the temporal activity patterns as observed from each grid cell. We start by normalizing $T$ and $P$ such that the total number of users at a given time on a given day is equal to $1$
\begin{align}
\hat{T}_{g_0,w,h} & =  \frac{\displaystyle T_{g_0,w,h}}{\sum_{g=1}^n \displaystyle T_{g,w,h}}, \\
\hat{P}_{g_0,w,h} & =  \frac{\displaystyle P_{g_0,w,h}}{\sum_{g=1}^n \displaystyle P_{g,w,h}} .
\label{chap}
\end{align}
This normalization allows for a direct comparison between sources with different absolute user's activity. For a given grid cell $g=g_0$, we defined the temporal distribution of users $\hat{P}_{g_0}$ as the concatenation of the temporal distribution of users associated with each day group. For each grid cell we obtained a temporal distribution of users represented by a vector of length $96$ corresponding to the $4 \times 24$ hours. 

After removing cells with zero temporal distribution, cells of common temporal profies were found using the ascending hierarchical clustering (AHC) method. The average linkage clustering and the Pearson correlation coefficient were taken as agglomeration method and similarity metric, respectively \cite{Hastie2009}. We have also implemented the k-means algorithm for extracting clusters but better silhouette index values were obtained with the AHC algorithm (see details in Figure S3 in Appendix). To choose the number of clusters, we used the average silhouette index $\bar{S}$ \cite{Rousseeuw1987}. For each cell $g$, we can compute $a(g)$ the average dissimilarity of $g$ (based on the Pearson correlation coefficient in our case) with all the other cells in the cluster to which $g$ belongs. In the same way, we can compute the average dissimilarities of $g$ to the other clusters and define $b(g)$ as the lowest average dissimilarity among them. Using these two quantities, we compute the silhouette index $s(g)$ defined as
\begin{equation}
s(g)=\frac{b(g)-a(g)}{max\{a(g),b(g)\}} ,
\label{sg}
\end{equation}
which measures how well clustered $g$ is. This measure is comprised between $-1$ for a very poor clustering quality and $1$ for an appropriately clustered $g$. We choose the number of clusters that maximize the average silhouette index over all the grid cells $\bar{S}=\sum_{g=1}^n s(g)/n$.

\begin{figure*}
\centering
\includegraphics[width=13cm]{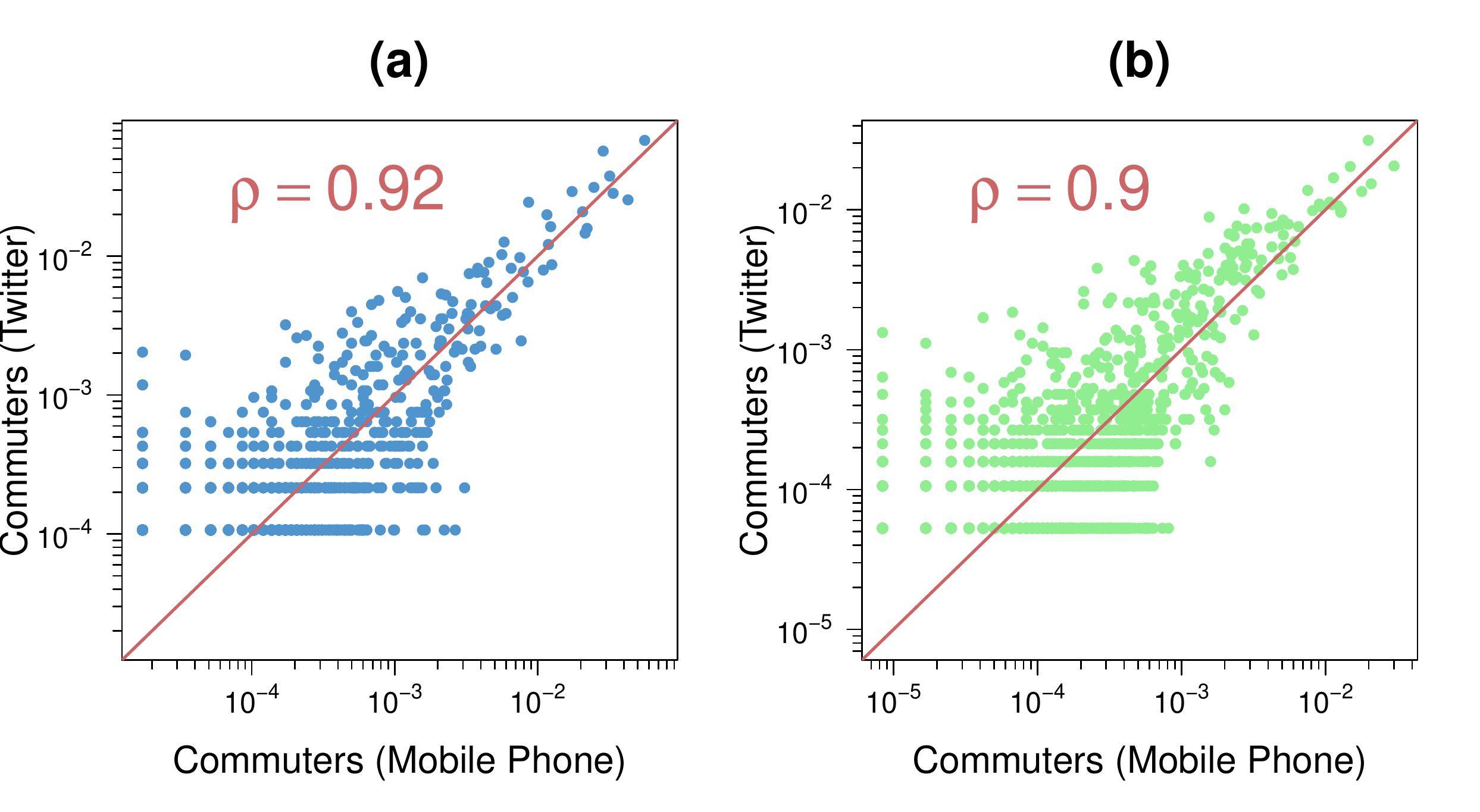}
\caption{Comparison between the non-zero flows obtained with the Twitter dataset and the mobile phone dataset (the values have been normalized by the total number of commuters for both OD tables). The points are scatter plot for each pair of grid cells. The red line represents the $x=y$ line. (a) Barcelona. (b) Madrid. In both cases $l=2$ $km$. \label{OD}}
\end{figure*}

For the mobile phone data, three clusters were found with an average silhouette index equal to $0.38$ for Barcelona and to $0.43$ for Madrid. The three temporal distribution patterns of mobile phone users are shown in Figure \ref{Patterns} for Barcelona. These three clusters can be associated with the following land uses:
\begin{itemize}
   \item \textbf{Business:} this cluster is characterized by a higher activity during the weekdays than the weekend days. In Figure \ref{Patterns}a, we observe that the activity takes place between $6$ am and $3$ pm with a higher activity during the morning. 
   \item \textbf{Residential:} this cluster is characterized by a higher activity during the weekend days than during the weekdays. Figure \ref{Patterns}c shows that the activity is almost constant from $9$ am during the weekend days. During the weekdays we observe two peaks, the first one between $7$ am and $8$ am and the second one during the evening. 
	 \item \textbf{Nightlife:} this cluster is characterized by a high activity during the night especially the weekend (Figure \ref{Patterns}e). 
\end{itemize}
It is remarkable to note that we obtain the same three patterns for Madrid and that these patterns are robust for different values of the scale parameter $l$ (see details in Figure S4, S5 and S6 in Appendix).

For Twitter data, considering a number of clusters smaller than 10, silhouette index values lower than 0.1 are obtained for both case studies. These low values mean that no clusters have been detected in the data probably because the Twitter data are too noisy. A way to bypass this limitation is to check if, for both data sources, the same patterns are obtained considering the different clusters obtained with the mobile phone data. To do so the temporal distribution patterns of Twitter users associated with the three clusters obtained with the mobile phone data are computed. We note in Figure \ref{Patterns} that for Barcelona the temporal distribution patterns obtained with the Twitter data are very similar to those obtained with the mobile phone data. We obtain the same correlation for Madrid and for different values of the scale $l$ (see details in Figure S4, S5 and S6 in Appendix).

\subsection{Users' mobility}
In this section, we study the similarity between the OD matrices extracted from Twitter and cell phone data. As it involves a change of spatial resolution needing extra attention, the comparison with the census is relegated to a coming section. We are able to infer for the metropolitan areas of Barcelona and Madrid the number of individuals living in the cell $i$ and working in the cell $j$. Figure \ref{OD} shows a scattered plot with the comparison between the flows obtained in the OD matrices for links present in both networks. In order to compare the two networks, the values have been normalized by the total number of commuters.

\begin{figure*}
\centering
\includegraphics[width=13cm]{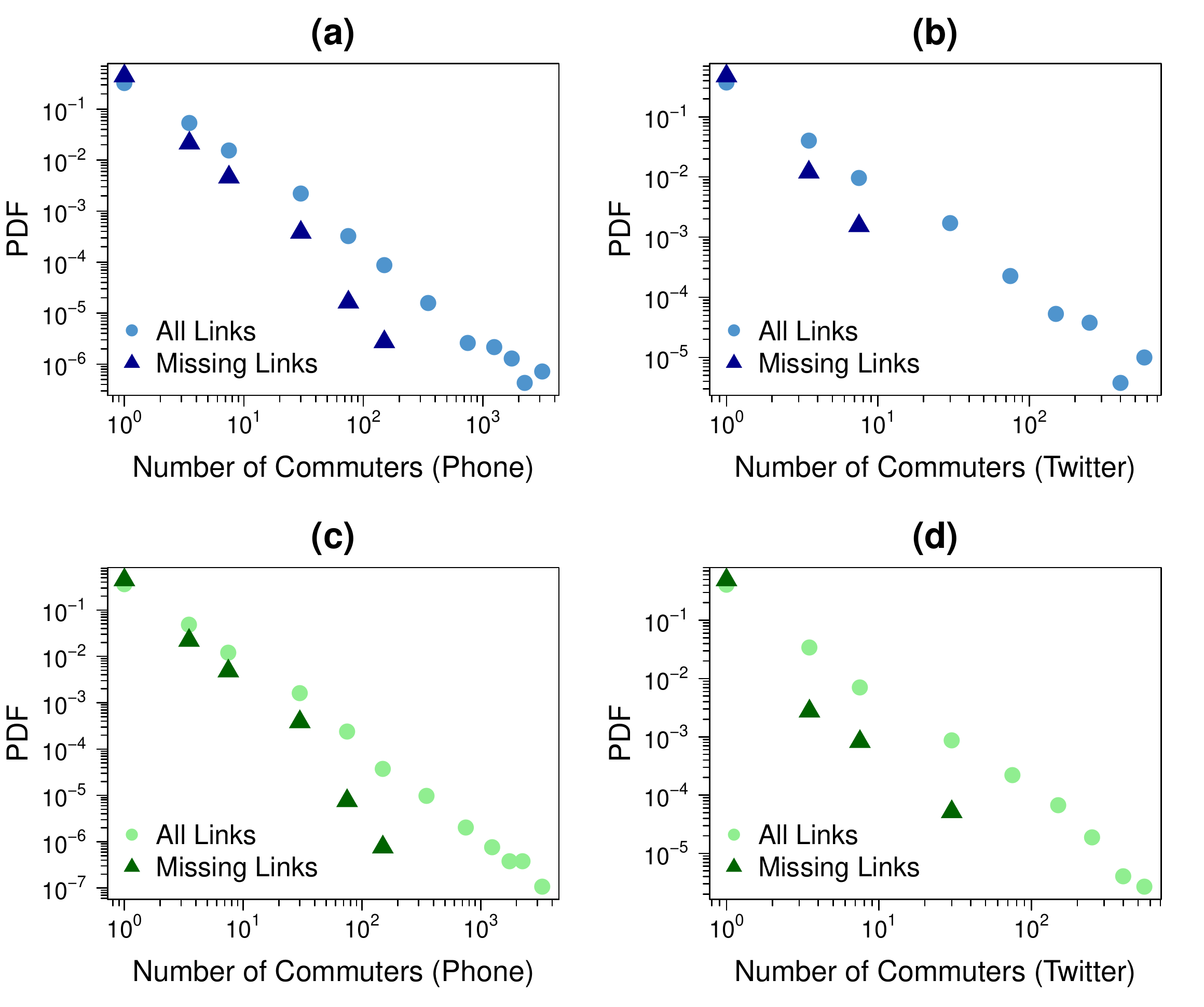}
\caption{Probability density function of the weights considering all the links (points) and the missing links (triangles). (a) Barcelona and cell phone data. (b) Barcelona and Twitter data. (c) Madrid and cell phone data. (d) Madrid and Twitter data. In both cases $l=2$ $km$. \label{miss}}
\end{figure*}

The overall agreement is good, the Pearson correlation coefficient is around $\rho \approx 0.9$. This coefficient measures the strength of the linear relationship between the normalized flows extracted from both networks, including the zero flows (i.e. flows with zero commuters). However, a high correlation value is not sufficient to assess the goodness of fit. Since we are estimating the fraction of commuters on each link, the values obtained from Twitter and the cell phone data should be ideally not only linearly related but the same. That is, if $y$ if the estimated fraction of mobile phone users on a connection and $x$ the estimated Twitter users on the same link, there should be not only a linear relation, which involves a high Pearson correlation, but also $y = x$. It is, therefore, important to verify that the slope of the relationship is equal to one. To do so, the coefficients of determination $R^2$ are computed to measure how well the scatterplot if fitted by the curve $y = x$. Since there is no particular preference for any set of data as $x$ or $y$, two coefficients $R^2$ can be measured, one using Twitter data as the independent variable $x$ and another using cell phone data. Note that if the slope of the relationship is strictly equal to one the two $R^2$ must be equal to the square of the correlation coefficient, we obtain a value around $R^2 = 0.85$ for Barcelona and around $0.81$ for Madrid. The slope of the best fit is in both cases very close to one.

The dispersion in the points is higher in low flow links. This can be explained by the stronger role played by the statistical fluctuations in low traffic numbers. Moreover, if we increase the resolution by using a value of $l$ equal to $1$ $km$, the Pearson correlation coefficient remains high with a value around $0.8$ (see details in Figure S7 in Appendix). The extreme situation of these fluctuations occurs when a link is present in one network and it has zero flow in the other (missing links). On average $90 \%$ of these links have a number of commuters equal to one in the network in which they are present. This shows that the two networks are not only inferring the same mobility patterns, but that the information left outside in the cross-check corresponds to the weakest links in the system. In order to assess the relevance of the missing links, the weight distributions of these links is displayed in Figure \ref{miss} for all the networks and case studies. As a comparison line, the weight distribution of all the links are also shown in the different panels. In all cases, the missing links have flows at least one order of magnitude, sometimes two orders, lower than the strongest links in the corresponding networks. To be more precise, the strongest flow of the missing links is, depending on the case, between $25$ and $464$ times lower than the highest weight of all the links. Furthermore, the average weight of the missing links is between $4$ and $9$ times lower than that obtained over all the links. Most of the missing links are therefore negligible in the general network picture. 

\begin{figure*}
\centering
\includegraphics[width=13cm]{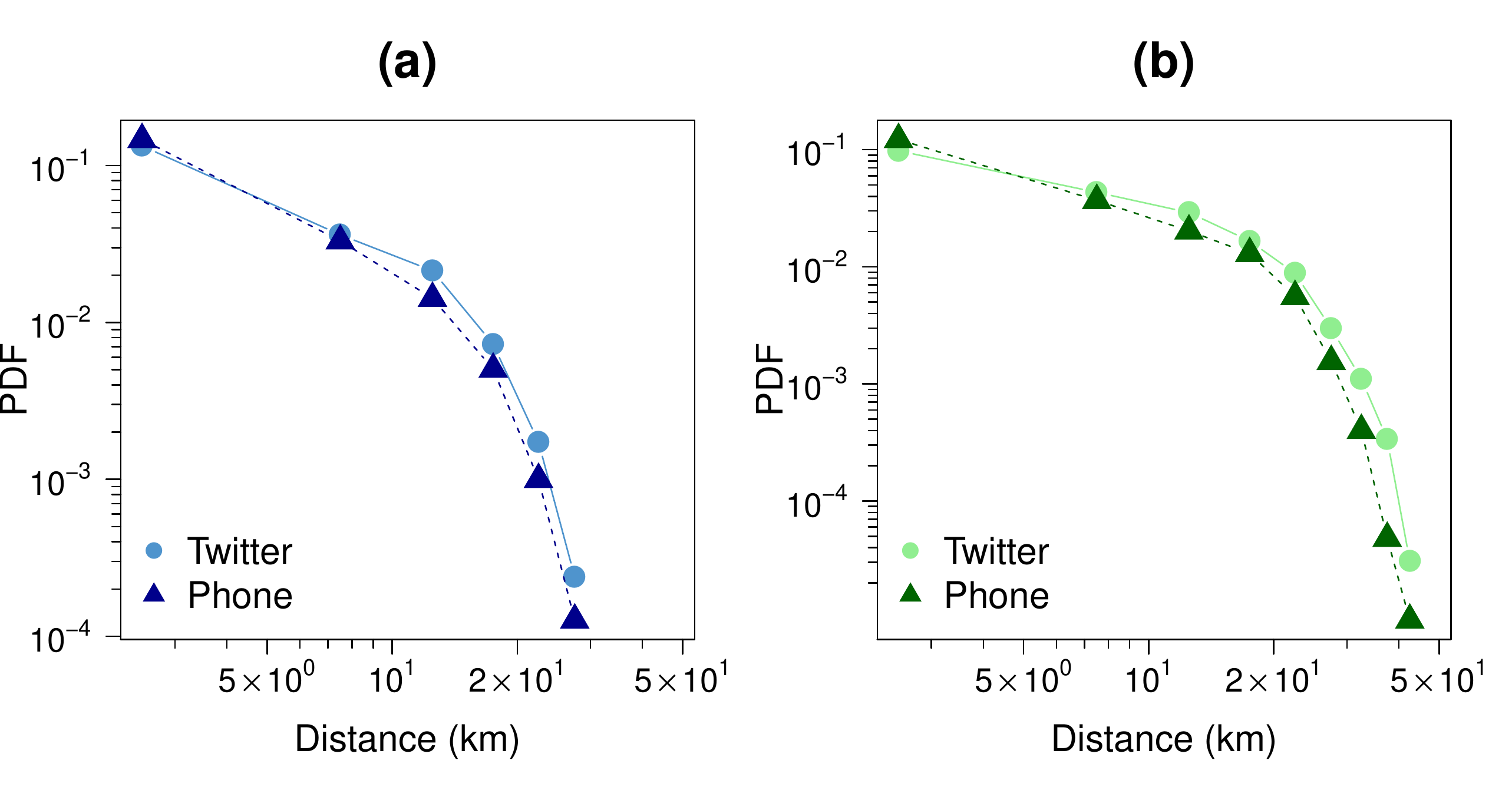}
\caption{Commuting distance distribution obtained with both datasets. We only consider individuals living and working in two different grid cells. The circles represent the Twitter data and the triangles the mobile phone data. (a) Barcelona. (b) Madrid. In both cases $l=2$ $km$. \label{CDD}}
\end{figure*}

With the aim of going a little further, we analyze and compare next the distance distribution for the trips obtained from both datasets. The geographical distance along each link in the OD matrices is calculated and the number of people traveling in the links is taken into account to evaluate the travel-length distribution. Figure \ref{CDD} shows these distributions for each network. Strong similarity between the two distributions can be observed in the two cities considered.

\subsection{Census, Twitter and cell phone}

As a final cross-validation, we compare the OD matrices estimated in workdays from Twitter and cell phone data to those extracted from the $2011$ census in Barcelona and Madrid. The census data is at the municipal level, which implies that to be able to perform the comparative analysis the geographical scale of both Twitter and phone data must be modified. To this end, the GIS files with the border of each municipality were used, instead of the grid, to compute the OD matrices from Twitter and cell phone data. Figure \ref{OD2} shows a scattered plot with the comparison between the flows obtained with the three networks. A good agreement between the three datasets is obtained with a Pearson correlation coefficient around $\rho \approx 0.99$. As mentioned previously, the correlation coefficient is not sufficient to assess the goodness of fit between the two networks. Thus, we have also computed two coefficients of determination $R^2$ for each one of the three relationships to measure how well the line $x=y$ approximates the scatter plots. For the two first relationships, the comparison between the Twitter and the mobile phone and the comparison between the mobile phone and the census OD tables,  we obtain $R^2$ values higher than $0.95$. For the last relationship (Twitter vs census), two different $R^2$ values are obtained because the best fit slope of the scatter plot is not strictly equal to one (0.85). The first $R^2$ value, which measure how well the normalized flows obtained in the Twitter's OD matrix approximate the normalized flows obtained in the census's OD matrix, is equal to $0.8$ and the second value, which assess the quality of the opposite relationship, is equal to $0.9$. A better result is instead obtained for Madrid with a Pearson correlation coefficient around $0.99$ and coefficients of determination higher than $0.97$ (see details in Figure S8 in Appendix).

\section{DISCUSSION}

In summary, we have analyzed mobility in urban areas extracted from different sources: cell phones, Twitter and census. The nature of the three data sources is very different, as also is the resolution scales in which the mobility information is recovered. For this reason, the aim of this work has been to run a thorough comparison between the information collected at different spatial and temporal scales. The first aspect considered refers to the population concentration in different parts of the cities. This point is of great importance in the analysis and planning of urban environments, including the design of new services or of contingency plans in case of disasters. Our results show that both Twitter and cell phone data produce similar density patterns both in space and time, with a Pearson correlation close to $0.9$ in the two cities analyzed. The second aspect considered has been the temporal distribution of individuals which allow us to determine the type of activity that are most common in specific urban areas. We show that similar temporal distribution patterns can be extracted from both Twitter and cell phone datasets. The last question studied has been the extraction of mobility networks in the shape of Origin-Destination commuting matrices. We observe that at high spatial resolution, in grid cells with sides of $1$ or $2$ $km$, the networks obtained with both cell phones and Twitter are comparable. Of course, the integration time needed for Twitter is higher in order to obtain similar results. Twitter data can run in serious problems too if instead of recurrent mobility the focus is on shorter term mobility, but this point falls beyond the scope of this work. Finally, the comparison with census data is also acceptable: both Twitter and cell phone data reproduce the commuting networks at the municipal scale from an overall perspective. Still and although good on average, the agreement between the three different datasets is broken in some particular connections that deviate from the diagonal in our scatterplots. This can be explained by the fact that the datasets come from different sources, were collected in different years and may have different biases and level of representativeness. For example, Twitter is supposed to be used more by younger people. The explanation of these deviations and whether they are just stochastic fluctuations or follow some rationale could be an interesting avenue for further research.

\begin{figure*}
\centering
\includegraphics[width=15cm]{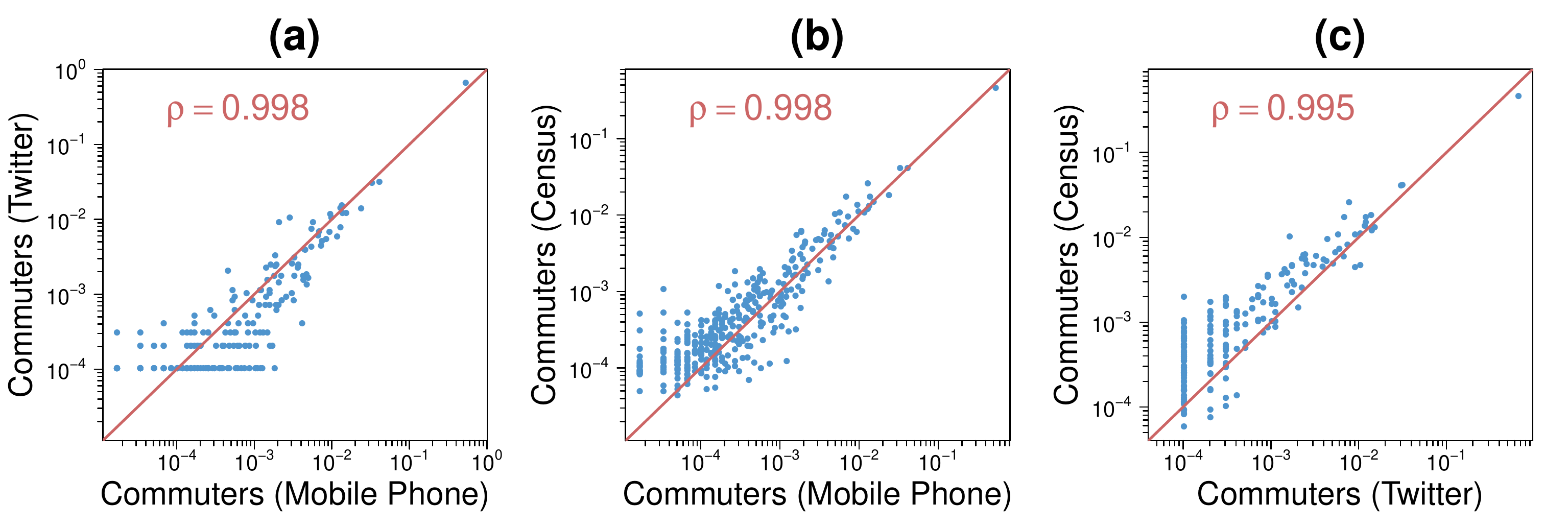}
\caption{Comparison between the non-zero flows obtained with the three datasets for the Barcelona's case study (the values have been normalized by the total number of commuters for both OD tables). Blue points are scatter plot for each pair of municipalities. The red line represents the $x=y$ line. (a) Twitter and mobile phone. (b) Census and mobile phone. (c) Census and Twitter. \label{OD2}}
\end{figure*} 

These results set a basis for the reliability of previous works basing their analysis on single datasets. Similarly, the door to extract conclusions from data coming from a single data source (due to convenience of facility of access) is open as long as the spatio-temporal scales tested here are respected.

\section{ACKNOWLEDGEMENTS}
Partial financial support has been received from the Spanish Ministry of Economy (MINECO) and FEDER (EU) under projects MODASS (FIS2011-24785) and INTENSE\@COSYP (FIS2012-30634),  and from the EU Commission through projects EUNOIA, LASAGNE and INSIGHT. ML acknowledges funding from the Conselleria d'Educaci\'o, Cultura i Universitats of the Government of the Balearic Islands and JJR from the Ram\'on y Cajal program of MINECO.

\newpage
\clearpage
\newpage

\makeatletter
\renewcommand{\fnum@figure}{\small\textbf{\figurename~S\thefigure}}
\makeatother
\setcounter{figure}{0}

\setcounter{equation}{0}

\section*{APPENDIX}

\subsection*{Mobile phone data pre-processing}

\subsubsection*{Outliers detection}

\begin{figure*}
		\includegraphics[width=16cm]{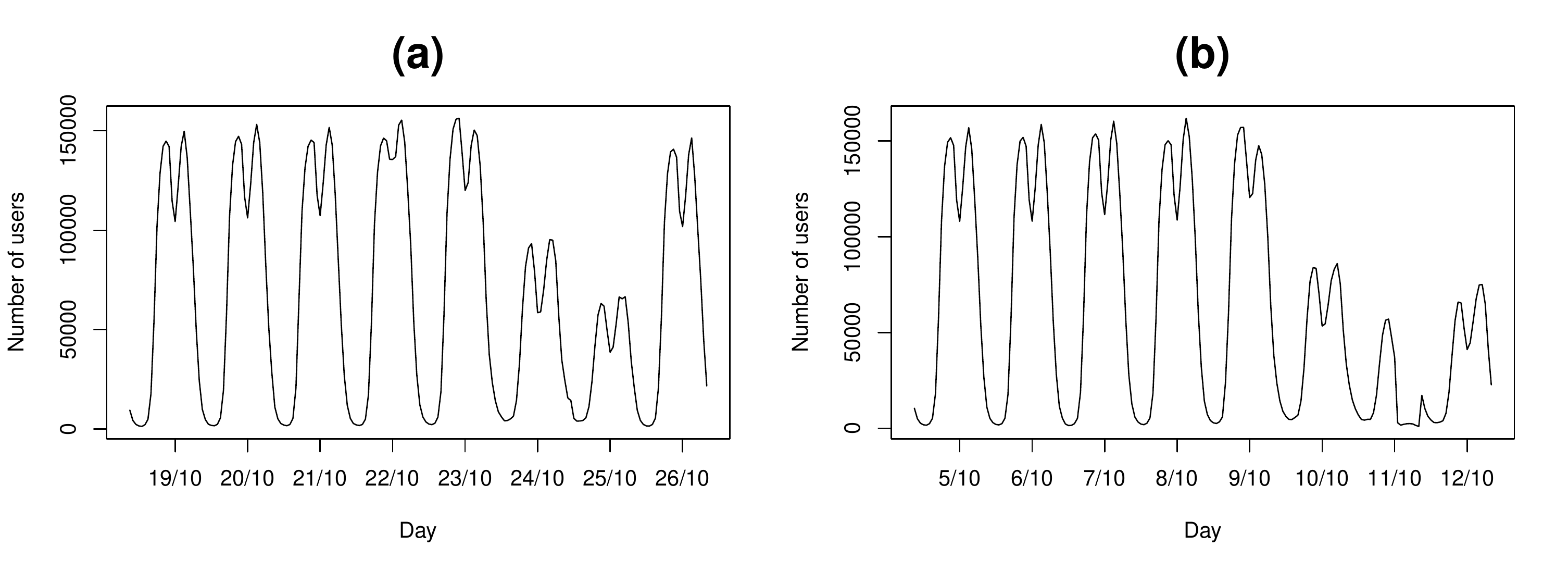}
		\caption{Temporal distribution of the mobile phone users for the metropolitan area of Barcelona. (a) From 19/10/2009 to 25/10/2009, eight days without outlier days. (b)  From 05/10/2009 to 12/10/2009, eight days with two outlier days (11/10/2009 and 12/10/2009). \label{S1}}
\end{figure*}

For both datasets we need to identify the outlier days to remove them from the data base. There are two types of outlier days, the special days (for example the National day) and the day for which we do not have the data for few hours. For example, for the metropolitan area of Barcelona, we can observe in Figure S\ref{S1}a eight days (from Monday to Monday) without outliers and in Figure S\ref{S1}b eight days with two outliers, Sunday, October 11$^{\mbox{th}}$  2009 for which we do not have the data from 5PM to 11PM and Monday, October 12$^{\mbox{th}}$  2009 the Spain's National Day.  

\subsubsection*{Voronoi cells}

\begin{figure*}
    \centering
		\includegraphics[width=16cm]{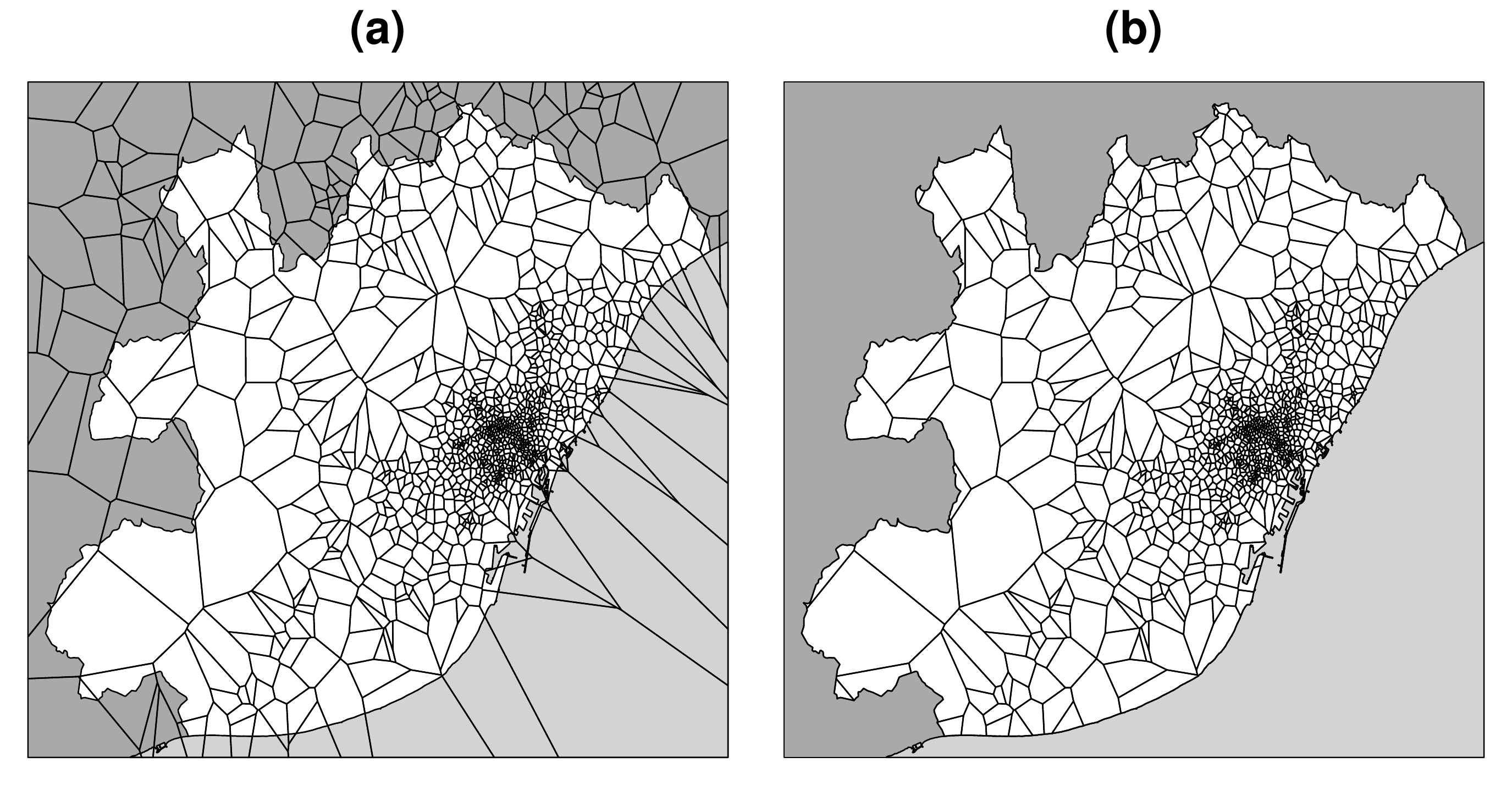}
		\caption{Map of the metropolitan area of Barcelona. The white area represents the metropolitan area, the dark grey area represents territory surrounding the metropolitan area and the gray area the sea. (a) Voronoi cells. (b) Intersection between the Voronoi cells and the metropolitan area.\label{S2}}
\end{figure*}

We remove the BTSs with zero mobile phone users and we compute the Voronoi cells associated with each BTSs of the metropolitan area (hereafter called MA). We remark in Figure S\ref{S2}a that there are four types of Voronoi cells:

\begin{enumerate}
   \item The Voronoi cells contained in the MA.
	 \item The Voronoi cells between the MA and the territory outside the metropolitan area.
	 \item The Voronoi cells between the MA and the sea (noted S).
	 \item The Voronoi cells between the MA, the territory outside the metropolitan area and the sea.
\end{enumerate}

To compute the number of users associated with the intersections between the Voronoi cells and the MA we have to take into account these different types of Voronoi cells. Let $m$ be the number of Voronoi cells, $N_{v}$ the number of mobile phone users in the Voronoi cell $v$ and $A_{v}$ the area of the Voronoi cell $v$, $v \in |[1,m]|$. The number of users $N_{v\cap MA}$ in the intersection between $v$ and MA is given by the following equation:

\begin{equation}
	N_{v\cap MA}=N_v \left(\frac{\displaystyle A_{v\cap MA}}{A_v - A_{v\cap S}}\right)
	\label{vMA}
\end{equation}

We note in Equation \ref{vMA} that we remove the intersection of the Voronoi area with the sea, indeed, we assume that the number of users calling from the sea are negligeable. Now we consider the number of mobile phone users $N_v$ and the associated area $A_v$ of the Voronoi cells intersecting the MA (Figure S\ref{S2}b).

\vspace{1cm}

\subsection*{Origin-Destination matrices}

As mentioned in the section \textit{Extraction of commuting matrices} unlike the Twitter data we cannot directly extract an OD matrix between the grid cells with the mobile phone data because each users' home and work locations are identified by the Voronoi cells. Thus, we need a transition matrix $P$ to transform the BTS OD matrix $B$ into a grid OD matrix $G$. 

Let $m$ be the number of Voronoi cells and $n$ be the number of grid cells. Let $B$ be the OD matrix between BTSs where $B_{ij}$ is the number of commuters between the BTS $i$ and the BTS $j$. To transform the matrix $B$ into an OD matrix between grid cells $G$ we define the transition matrix $P$ where $P_{ij}$ is the area of the intersection between the grid cell $i$ and the BTS $j$. Then we normalize $P$ by column in order to consider a proportion of the BTSs areas instead of an absolut value, thus we obtain a new matrix $\hat{P}$ (Equation S\ref{pchap}). 

\begin{equation}
	\hat{P}_{ij}=\frac{\displaystyle P_{ij}}{\sum_{k=1}^m \displaystyle P_{kj}}
	\label{pchap}
\end{equation}

The OD matrix between the grid cells $G$ is given by a matrices multiplication given in the following equation:

\begin{equation}
	 G=P B P^t
	\label{OD3}
\end{equation}


\begin{figure*}
    \centering 
		\includegraphics[scale=0.65]{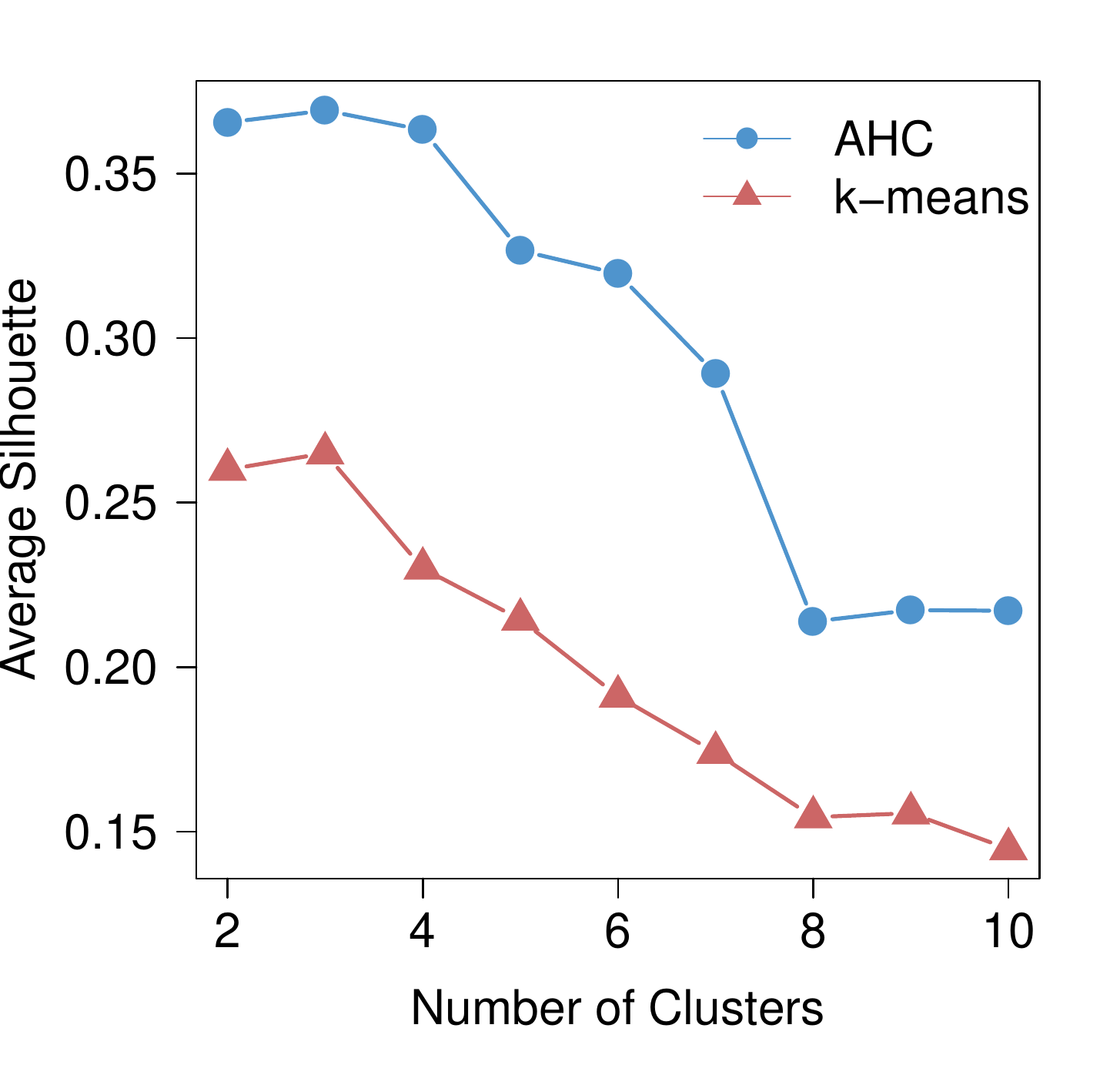}.
		\caption{Average Silhouette as a function of the number of clusters obtained with AHC (in blue) and k-means (in red).\label{FigS3}}
\end{figure*}

\begin{figure*}
    \centering 
		\includegraphics[scale=0.45]{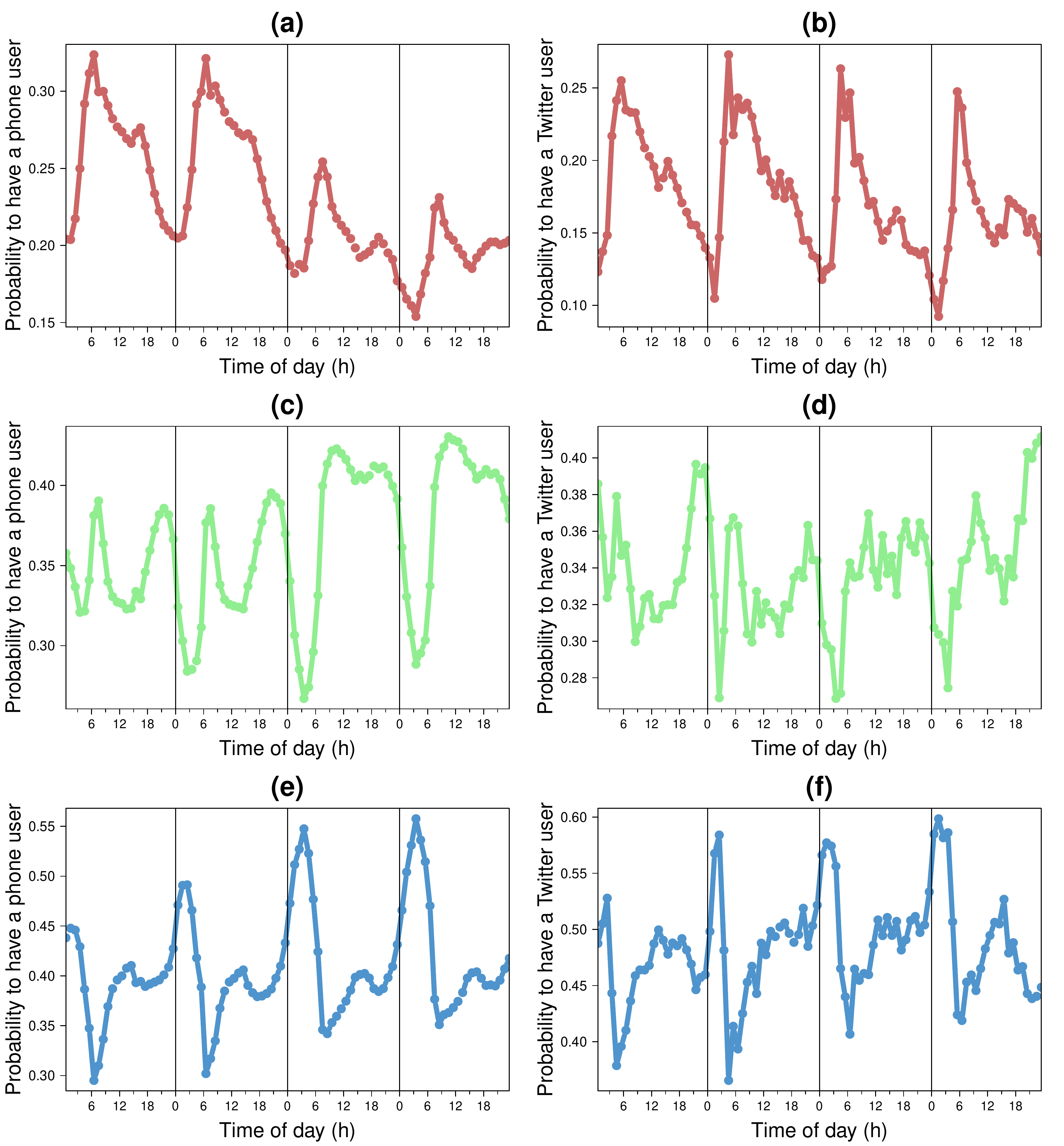}.
		\caption{Temporal distribution patterns for the metropolitan area of Madrid ($l=2$). (a), (c) and (e) Mobile phone activity; (b), (d) and (f) Twitter activity; (a) and (b) Business cluster; (c) and (d) Residential/leisure cluster; (e) and (f) Nightlife cluster.\label{FigS4}}
\end{figure*}

\begin{figure*}
    \centering 
		\includegraphics[scale=0.45]{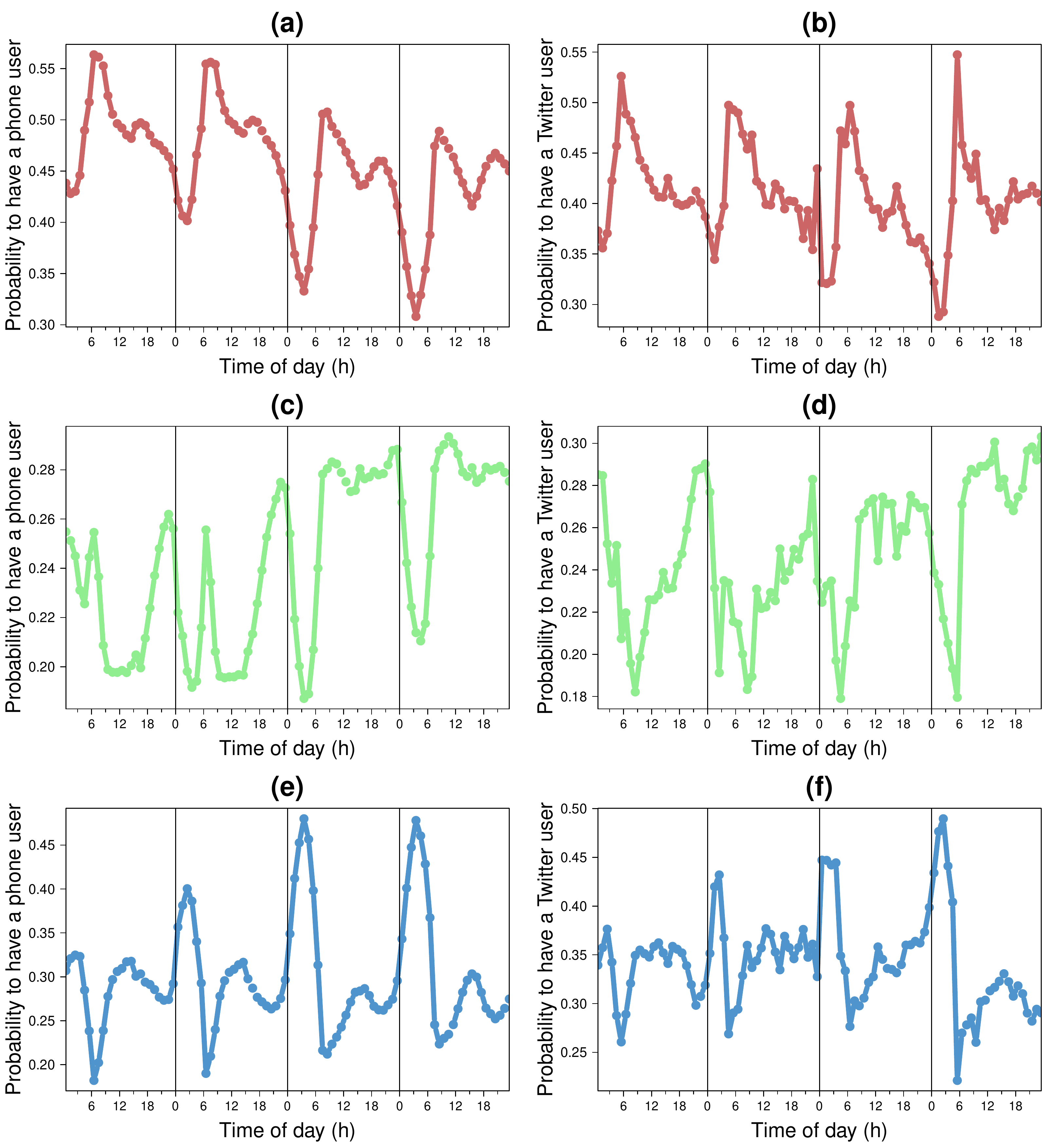}.
		\caption{Temporal distribution patterns for the metropolitan area of Barcelona ($l=1$). (a), (c) and (e) Mobile phone activity; (b), (d) and (f) Twitter activity; (a) and (b) Business cluster; (c) and (d) Residential/leisure cluster; (e) and (f) Nightlife cluster.\label{FigS5}}
\end{figure*}

\begin{figure*}
    \centering 
		\includegraphics[scale=0.45]{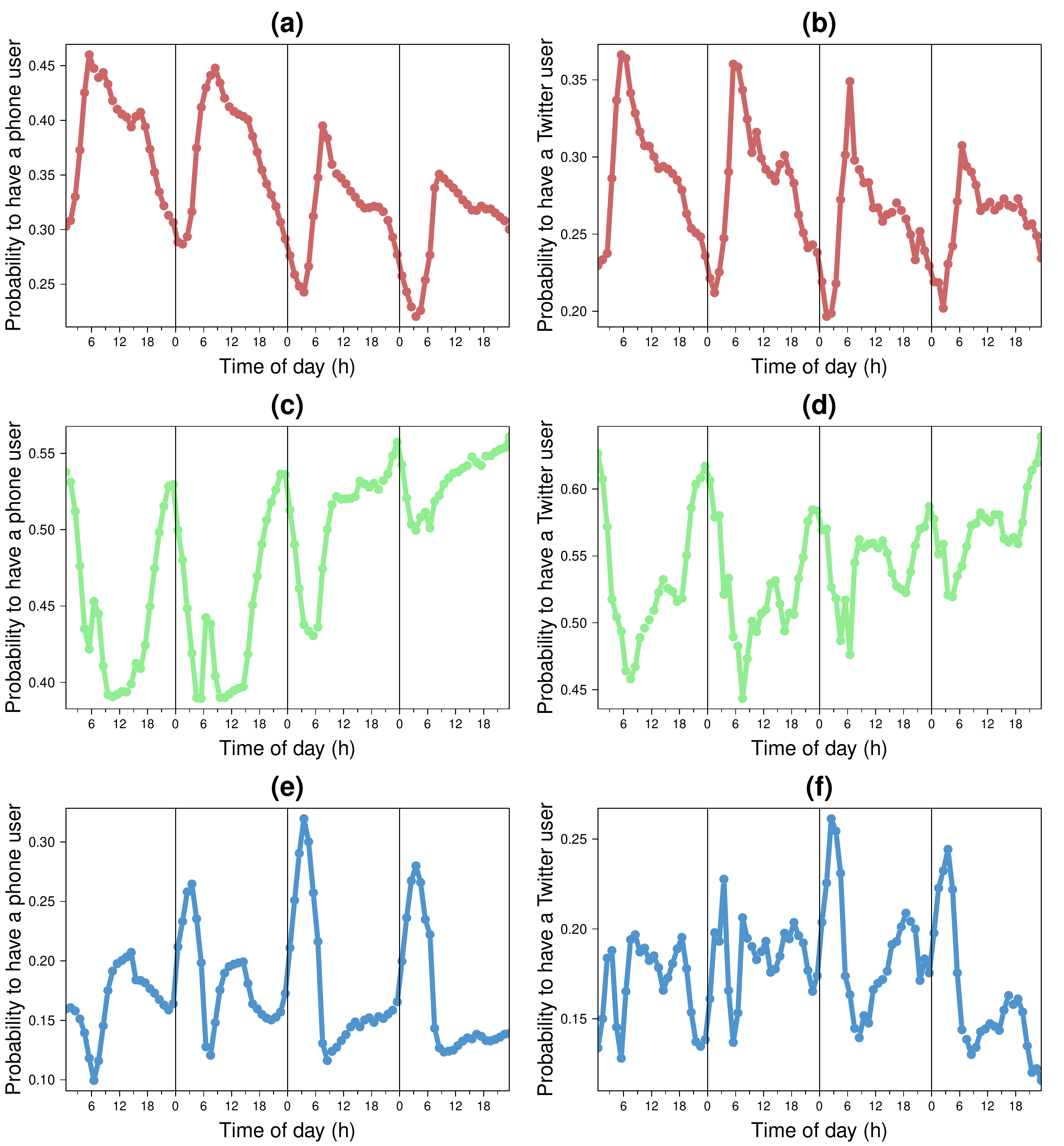}.
		\caption{Temporal distribution patterns for the metropolitan area of Madrid ($l=1$). (a), (c) and (e) Mobile phone activity; (b), (d) and (f) Twitter activity; (a) and (b) Business cluster; (c) and (d) Residential/leisure cluster; (e) and (f) Nightlife cluster.\label{FigS6}}
\end{figure*}

\begin{figure*}
\centering
\includegraphics[scale=0.6]{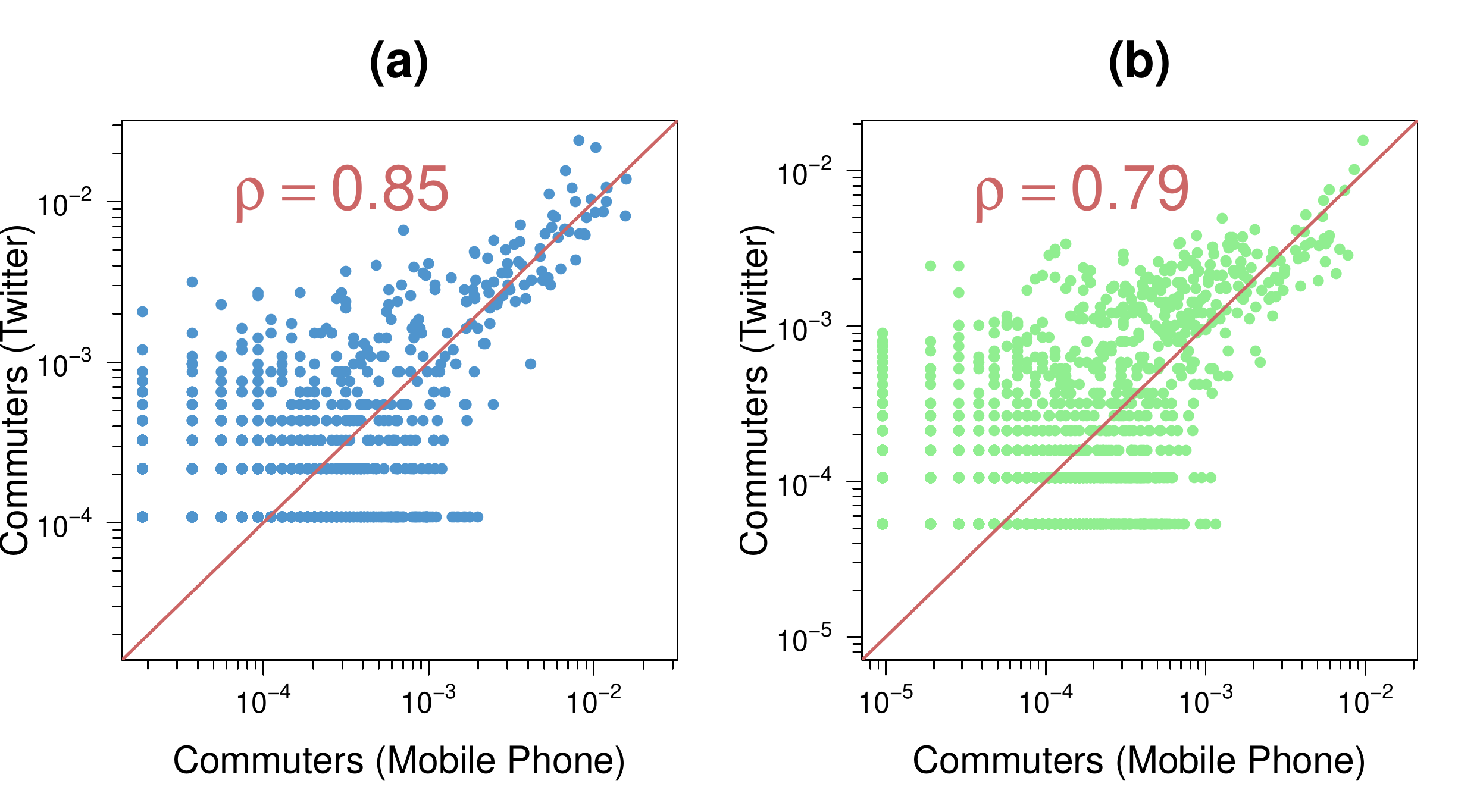}
\caption{Comparison between the non-zero flows obtained with the Twitter dataset and the mobile phone dataset (the values have been normalized by the total number of commuters for both OD tables). The points are scatter plot for each pair of grid cell. The red line represents the $x=y$ line. (a) Barcelona. (b) Madrid. In both cases $l=1$ $km$. \label{S7}}
\end{figure*}

\begin{figure*}
\centering
\includegraphics[width=\linewidth]{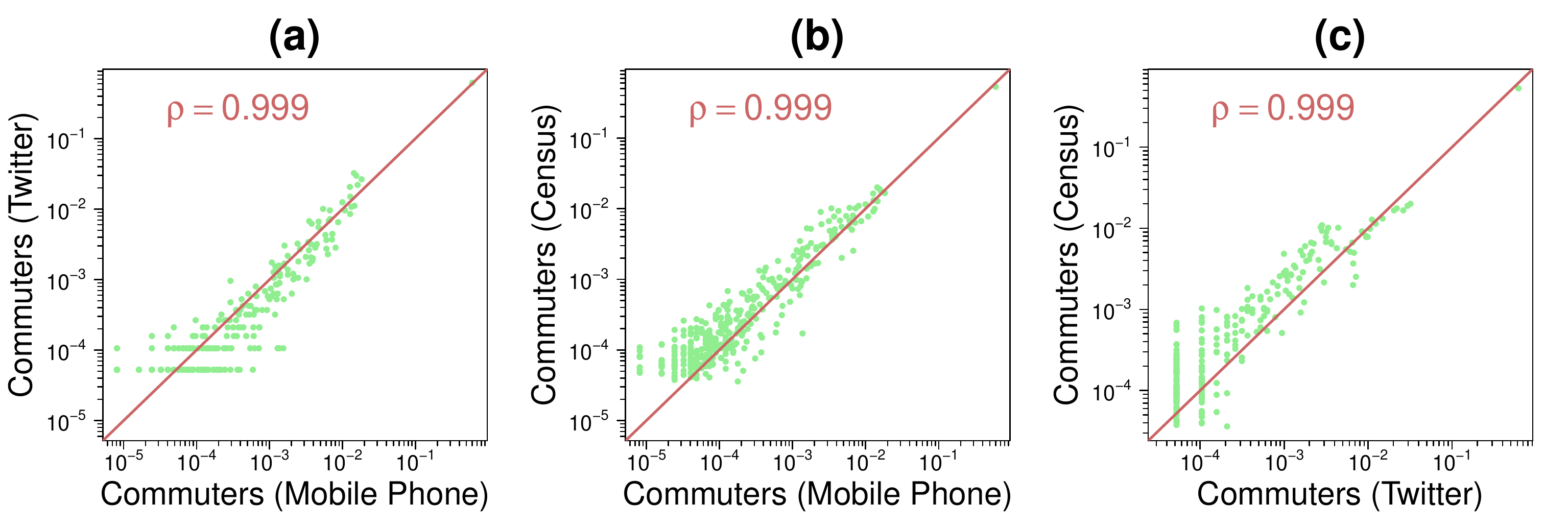}
\caption{Comparison between the non-zero flows obtained with the three datasets for the Madrid's case study (the values have been normalized by the total number of commuters for both OD tables). Green points are scatter plot for each pair of municipalities. The red line represents the $x=y$ line. (a) Twitter and mobile phone. (b) Census and mobile phone. (c) Census and Twitter. \label{OD2}}
\end{figure*}

\end{document}